\documentclass[twocolumn,showpacs,preprintnumbers,amsmath,amssymb,aps]{revtex4} 

\def\ie {i.e.} 

\begin{document}

\title{Conformally invariant wave equation for a symmetric second rank tensor (``spin-2'') in d-dimensional curved background. }

\author{J. Ben Achour$^1$}
\author{E. Huguet$^1$}
\author{J. Renaud$^2$}
\affiliation{$1$ - Universit\'e Paris Diderot-Paris 7, APC-Astroparticule et Cosmologie (UMR-CNRS 7164), 
Batiment Condorcet, 10 rue Alice Domon et L\'eonie Duquet, F-75205 Paris Cedex 13, France.  \\
$2$ - Universit\'e Paris-Est, APC-Astroparticule et Cosmologie (UMR-CNRS 7164), 
Batiment Condorcet, 10 rue Alice Domon et L\'eonie Duquet, F-75205 Paris Cedex 13, France.
} 
\email{benachou@apc.univ-paris7.fr, huguet@apc.univ-paris7.fr, jacques.renaud@u-pem.fr}

\date{\today}


\begin{abstract}
We build the general conformally invariant linear wave operator for a free, symmetric, second rank tensor field 
in a d-dimensional ($d\geqslant 2$) metric manifold,  and explicit the special case of maximally symmetric spaces. 
Under the assumptions made, this conformally invariant wave operator is 
unique.
The corresponding conformally invariant wave equation can be obtained from a 
Lagrangian which is explicitly given.  We discuss how our result compares to previous works, in particular we hope to 
clarify the situation between conflicting results. 
\end{abstract}

\maketitle

\section{Introduction}\label{Sec-Introduction}

In this paper we derive the most general conformally invariant linear (second order) wave operator for a free, symmetric, 
second order tensor field in a d-dimensional spacetime (l.h.s. of Eq. (\ref{EQ-GeneralFormCoeff}) with coefficients 
(\ref{EQ-ValeurCoeff})). We compare this result with previous works, some of them giving contradictory results.

The conformal transformation is of daily uses in many areas of theoretical physics \cite{kastrup}. The conformal 
invariance (see for instance \cite{wald}) or the conformal mapping, are frequently used as a mathematical tool in the
resolution of differential equations.  In particle physics it is often associated to masslessness and light-cone
propagation, although it is known for a while that this relation can be broken in curved spacetime or $d\neq4$ dimensions \cite{DeserNepo}.  
An important instance is given by the hypothetical graviton, which is expected to be a massless spin-2 particle despite the fact that 
the Einstein equations, and their linearized version, are not invariant under conformal transformations. This has been studied, in particular, by 
Grishchuk and Yudin \cite{GrishchukYudin1} in the fall of the seventies. The main consequence of the non-conformality of the linearized Einstein equations 
is that gravitational waves (and their possible quantum counterpart) can be created in early Universe as a result of the expansion \cite{Grishchuk1,Parker}.

Although the linearized Einstein equations are not conformally invariant, it is natural to wonder if there is a conformally invariant wave equation for
a symmetric second order tensor field. Beside its intrinsic interest as a field equation, it could be also interesting to study how such an 
equation relates to other ``spin-2'' equations, in particular in conformally related spacetimes such as Robertson-Walker spacetimes. This has been 
already considered in \cite{GrishchukYudin1} under some rather stringent assumptions, namely: $\nabla_\mu h^\mu_{~\nu}  = 0$ and 
$h^\mu_{~\nu} \nabla_\mu \log \Omega  = 0$, 
where $h_{\mu\nu}$ is the ``spin-2'' field and $\Omega$ the conformal factor. 
Recently, a wave equation, in which these assumptions are relaxed, has been proposed by Faci \cite{Faci} as a byproduct of a method designed for 
building conformally invariant equations from Weyl geometry. 
In fact,  such a  wave equation, invariant under an {\it a priori} smaller set of symmetries, has already been obtained in $d$ dimensional 
spaces ($d>2$) by Nepomechie \cite{Nepomechie} in a work relating the low energy limit of Einstein and Weyl Gravity.  The case of 
conformally covariant operators acting on traceless symmetric tensors of arbitrary rank in $d$ dimensions ($d>2$) has been also considered
by Erdmenger and Osborn \cite{ErdmengerOsborn}.

A first attempt to use some of these results led us to question their reliability. Afterwards, a comparison between all these 
works showed us they disagree the ones with the others.
Therefore, we find it useful to clarify the situation in order to have a reliable general conformal wave equation for a 
symmetric second order tensor field. Consequently, despite the existence of these works, the present paper is devoted to the derivation 
of a Weyl invariant wave equation for a symmetric second rank tensor, or more precisely a Weyl invariant operator from which the invariant equation follows.

We find that this operator (l.h.s. of Eq. (\ref{EQ-GeneralFormCoeff}) with coefficients 
(\ref{EQ-ValeurCoeff})) is basically that found earlier by Nepomechie \cite{Nepomechie}. Our work shows in addition that this operator
is the most general and the only possibility under the assumptions made,  it makes
explicit it's Weyl invariance, and extends it to the case of dimension $d=2$.  
This operator is found to agree with that of Grishchuk and Yudin \cite{GrishchukYudin1} when the same constraints are applied. There is 
mild discrepancies, with the result of Faci \cite{Faci}, and a deeper disagreement with the equation deduced from the work of Erdmenger and Osborn 
\cite{ErdmengerOsborn}. In particular, this last work seems to provide an example of an invariant Lagrangian supplemented with an 
invariant constraint which leads to non-invariant Euler equations. We explain this somehow unexpected situation.

Finally, let us point out that since no restrictions are applied to the second rank symmetric tensor field we consider, it is not a 
true spin-$2$ field (although this misnomer is sometimes encountered in the literature). In particular, it is worth noting that it contains ghosts.

The paper is organized as follows. In Sec. \ref{Sec-Equation} we derive the conformal wave equation for dimensions
$d > 2$ including the specialization to maximally symmetric spaces.  The special case of $d=2$ is addressed in Sec. \ref{Sec-dim2}.
The Sec. \ref{Sec-Compar} is devoted to the comparison with the earlier results mentioned above.  We conclude in Sec. \ref{Sec-conl}. Since 
a large part of this work rests on explicit calculations we give for convenience some details and hopefully reliable reference formulas, 
including the adopted conventions, in appendices. 

\section{Conformally invariant wave equation}\label{Sec-Equation}
Let us consider a conformal mapping between two d-dimensional metric manifolds $(\mathcal{M},g)$ and ($\mathcal{M},\overline{g})$, the two metrics $g$ 
and $\overline{g}$ being related through $\overline{g}_{\mu\nu} = \Omega^2(x) g_{\mu\nu}$. The conformal factor $\Omega$ being a positive well behaved
function defined on $\mathcal{M}$. As usual we denote by a bar the transformed quantities. We are looking for a conformally invariant wave 
equation for a free second rank symmetric tensor of conformal weight $r$.
By wave equation we mean a differential second order homogeneous equation in which the coefficient of  the term 
$\square h_{\mu\nu} = g^{\alpha\beta}\nabla_\alpha\nabla_\beta  h_{\mu\nu}$ is one and such that its other coefficients can contain 
exclusively the polynomial 
combinations of the metric tensor, the Riemann tensor and their contractions: 
the Ricci tensor and the Ricci scalar.

Under the  above hypothesis, the most general conformally invariant equation can be obtained with the same approach 
than that of Grishchuk and Yudin \cite{GrishchukYudin1}. 
Before reminding the main steps of this derivation, let us remark that the invariant object built is indeed a conformally 
invariant operator, that is an operator $\mathcal{E}$ acting on a field $\psi$ with some conformal weight $r$, \ie $\overline{\psi} = \Omega^r\psi$, 
such that under the conformal 
mapping $\overline{g}_{\mu\nu} = \Omega^2(x) g_{\mu\nu}$ one has 
$\overline{\mathcal{E}}\, \overline{\psi} = \Omega^q \mathcal{E}\psi$, $q$ being a conformal weight, $\overline{\mathcal{E}}$ having the same 
form as $\mathcal{E}$. For such an
operator the corresponding equation $ \mathcal{E}\psi = 0$ is conformally invariant, that is having a solution of the equation is equivalent 
to having a solution of its transform $\overline{\mathcal{E}}\, \overline{\psi}$ = 0. Note that the converse is not true in general.

Let us turn now to the derivation. The general form for a covariant equation of a 
second rank tensor reads:
\begin{equation}\label{EQ-GeneralForm}
\left(
\overline{F}_{\mu\nu}^{~~~\sigma\alpha\beta\rho} \overline{\nabla}_\sigma \overline{\nabla}_\rho  + 
\overline{P}_{\mu\nu}^{~~\alpha\beta\rho} \overline{\nabla}_\rho  +
\overline{U}_{~\mu\nu~}^{\alpha~~\beta} \right) \overline{h}_{\alpha\beta}=0
\end{equation}
where, taking into account the assumption made above, the tensors $F,P,U$ are polynomial combinations of the metric $g_{\mu\nu}$ the Riemann 
tensor $R_{\mu\alpha\beta\nu}$ and their contractions. Since no tensor with an odd number of indices can be built from such a combination we have $P=0$. 
Now, Eq. (\ref{EQ-GeneralForm}) is assumed to be symmetric in its free index $\mu\nu$. In addition, the identity
\begin{equation}\label{EQ-Commutateur}
\left[\nabla_{\mu},\nabla_{\nu}\right] h_{\alpha\beta} = R^\epsilon_{~\alpha\mu\nu} h_{\epsilon\beta} + R^\epsilon_{~\beta\mu\nu} h_{\alpha\epsilon},
\end{equation}
allows us to keep the symmetric part in $\rho\sigma$ of $(\overline{\nabla}_\sigma \overline{\nabla}_\rho)\overline{h}_{\alpha\beta}$ in the $\overline{F}$ term 
and to include its antisymmetric part into the $\overline{U}$ term. Finally, $\overline{F}_{\mu\nu}^{~~~\sigma\alpha\beta\rho}$ is symmetric in $\mu\nu$, 
$\alpha\beta$ (since $h_{\alpha\beta}$ is symmetric), and $\rho \sigma$; the tensor $\overline{U}_{~\mu\nu~}^{\alpha~~\beta}$ is symmetric in 
$\mu\nu$ and $\alpha\beta$. 

Under the conformal transformation $\overline{g}_{\mu\nu} = \Omega^2(x) g_{\mu\nu}$, the l.h.s. of Eq. (\ref{EQ-GeneralForm}), in order to be 
conformally invariant, 
has to transform into itself (without the bar) times a factor $\Omega^q$, $q$ being a conformal weight. Since we assume the presence of the term 
$\overline{\square}\,\overline{h}_{\mu\nu}$, this determines the value of $q$. The conformal transformation of $\overline{\square}\,\overline{h}_{\mu\nu}$ is
given in appendix \ref{AP-DetailWeylTransf}, one has $q=(r-2)$. For consistency, all other terms appearing in Eq. (\ref{EQ-GeneralForm}) have to transform 
with the same overall factor. 
This condition combined with the considerations of symmetry of the above paragraph allows us to determine all the terms contained in $\overline{F}$ 
and~$\overline{U}$. 

Without additional assumption about $h_{\mu\nu}$, the terms appearing 
in $\overline{F}$ are $\overline{\delta}_\mu^\sigma\overline{\delta}_\nu^\rho \overline{g}^{\alpha\beta}$,
$\overline{g}_{\mu\nu}\,\overline{g}^{\sigma\rho}\,\overline{g}^{\alpha\beta}$, and all the terms symmetrized in $\mu\nu$, $\sigma\rho$ 
and $\alpha\beta$. In the same way, the tensor $\overline{U}$ is built on the terms $\overline{g}_{\mu\nu}\overline{R}^{\alpha\beta}$, 
$\overline{R}_{\mu\nu}\,\overline{g}^{\alpha\beta}$, $\overline{R}\,\overline{g}_{\mu\nu}\overline{g}^{\alpha\beta}$, 
$\overline{R}\,\overline{\delta}_\mu^\alpha\overline{\delta}_\nu^\beta$, $\overline{R}^{\alpha~~~\beta}_{~\mu\nu~}$ and all the terms 
symmetrized in $\mu\nu$ and $\alpha\beta$. 
Collecting all the contributions to the terms $\overline{F}$ and $\overline{U}$ allows us to cast 
the Eq. (\ref{EQ-GeneralForm})
under the form
\begin{equation*}
\begin{split}
&\overline{\square}\,\overline{h}_{\mu\nu} + a \overline{R} \,\overline{h}_{\mu\nu} + 
b\left(\overline{R}_{\mu\alpha}\,\overline{h}^\alpha_{~\nu}+\overline{R}_{\nu\alpha}\,\overline{h}^\alpha_{~\mu}\right) 
+ c\overline{g}_{\mu\nu}\overline{R}_{\alpha\beta}\,\overline{h}^{\alpha\beta}\\
&+ e \left( \overline{\nabla}_\alpha \overline{\nabla}_\mu \,\overline{h}^\alpha_{~\nu}
+ \overline{\nabla}_\alpha \overline{\nabla}_\nu \,\overline{h}^\alpha_{~\mu}
+ \overline{\nabla}_\mu\overline{\nabla}_\alpha\,\overline{h}^\alpha_{~\nu}
+ \overline{\nabla}_\nu\overline{\nabla}_\alpha\,\overline{h}^\alpha_{~\mu}\right)\\
&+ p \overline{R}_{\mu\alpha\beta\nu}\overline{h}^{\alpha\beta} + f \overline{g}_{\mu\nu}\overline{\nabla}_\alpha\overline{\nabla}_\beta\,\overline{h}^{\alpha\beta} 
+ k \overline{g}_{\mu\nu} \overline{\square}\,\overline{h} \\
&+ l  \left( \overline{\nabla}_\mu \overline{\nabla}_\nu +\overline{\nabla}_\nu \overline{\nabla}_\mu \right) \overline{h}
+ m \overline{g}_{\mu\nu} \overline{R}\,\overline{h} + n \overline{R}_{\mu\nu}\overline{h} = 0,
\end{split}
\end{equation*}
in which $a,b,\ldots,m,n$ are unknown coefficients and $h := h^\alpha_\alpha$ is the trace of $h_{\mu\nu}$. Thanks to the relation
\begin{equation}\label{EQ-CommutDeriv}
\begin{split}
\left[\overline{\nabla}_\alpha, \overline{\nabla}_\mu\right] \overline{h}^\alpha_{~\nu} &= \overline{R}^\alpha_{~\sigma\mu\alpha}\overline{h}^\sigma_{~\nu}
-\overline{R}^\sigma_{~\nu\mu\alpha}\overline{h}^\alpha_{~\sigma},\\
&=-\overline{R}_{\mu\sigma}\overline{h}^\sigma_{~\nu} - \overline{R}_{\mu\alpha\sigma\nu}\overline{h}^{\alpha\sigma},
\end{split}
\end{equation}
and the symmetries of the Riemann tensor, the term whose coefficient is $e$ rewrites
\begin{equation*}
\begin{split}
&+2e\left( \overline{\nabla}_\mu\overline{\nabla}_\alpha\,\overline{h}^\alpha_{~\nu}
+ \overline{\nabla}_\nu\overline{\nabla}_\alpha\,\overline{h}^\alpha_{~\mu}\right)
-e \left( \overline{R}_{\mu\alpha}\overline{h}^\alpha_{~\nu} + \overline{R}_{\mu\alpha}\overline{h}^\alpha_{~\nu}\right)\\
&- 2e \overline{R}_{\mu\alpha\beta\nu}\overline{h}^{\alpha\beta}.
\end{split}
\end{equation*}
Then, after a redefinition of the coefficients $b$, $e$ and $p$ the equation becomes
\begin{equation}\label{EQ-GeneralFormCoeff}
\begin{split}
\overline{\square}\,\overline{h}_{\mu\nu} &+ a \overline{R}\, \overline{h}_{\mu\nu} + 
b\left(\overline{R}_{\mu\alpha}\,\overline{h}^\alpha_{~\nu}+\overline{R}_{\nu\alpha}\,\overline{h}^\alpha_{~\mu}\right)\\ 
&+ c\overline{g}_{\mu\nu}\overline{R}_{\alpha\beta}\,\overline{h}^{\alpha\beta}
+ e \left(\overline{\nabla}_\mu\overline{\nabla}_\alpha\,\overline{h}^\alpha_{~\nu}
+ \overline{\nabla}_\nu\overline{\nabla}_\alpha\,\overline{h}^\alpha_{~\mu}\right) \\
&+ p \overline{R}_{\mu\alpha\beta\nu}\overline{h}^{\alpha\beta} 
+ f \overline{g}_{\mu\nu}\overline{\nabla}_\alpha\overline{\nabla}_\beta\,\overline{h}^{\alpha\beta} 
+ k \overline{g}_{\mu\nu} \overline{\square}\,\overline{h} \\
&+ l  \left( \overline{\nabla}_\mu \overline{\nabla}_\nu +\overline{\nabla}_\nu \overline{\nabla}_\mu \right) \overline{h}
+ m \overline{g}_{\mu\nu} \overline{R}\,\overline{h} + n \overline{R}_{\mu\nu}\overline{h} = 0,
\end{split}
\end{equation}

A straightforward but rather lengthy calculation gives the conformal transformation of the above Eq. (\ref{EQ-GeneralFormCoeff}). The 
transformation of each term is detailed  in the Appendix~\ref{AP-DetailWeylTransf}. Under conformal mapping each term of 
Eq. (\ref{EQ-GeneralFormCoeff}) transforms as $\overline{X} = \Omega^{(r-2)}\left(X + \ldots\right)$. In order to be invariant under conformal 
transformations the additional terms have to cancel out. This translates into a system of equations for the coefficients. That system admits solutions 
which, for $d>2$, are parametrized by the dimension $d$ and the two coefficients $k$ and $p$ which are found to be unconstrained. The details of both the 
system and its resolution are given in Appendix~\ref{AP-DetailCalc}. The solutions read

\begin{alignat}{2}
a &= \frac{4(d+1)-d^2+4 p}{4 (d-1)(d-2)}, &b = -\frac{p+2}{d-2},\qquad\nonumber\\ 
c &= \frac{p d + 4}{(d-2) d}, &e= -\frac{4}{d+2},\qquad\nonumber\\ 
f &= \frac{8}{d (d+2)},  &l=\frac{f}{2},\label{EQ-ValeurCoeff}\qquad\\
r &= \frac{6-d}{2},   &n = c,\qquad\nonumber\\
m &= -\frac{8}{(d-1) (d^2-4)}  - &\frac{p}{(d-1) (d-2)}   -\frac{k}{4}\frac{d-2}{d-1}, \nonumber
\end{alignat}
where we recall that $r$ is the conformal weight of $h_{\mu\nu}$. The l.h.s. of Equation (\ref{EQ-GeneralFormCoeff})
with the above values of the coefficients $a, b, \ldots,n$ is the only conformally invariant wave operator in a general 
spacetime of dimension $d>2$. Note that $h_{\mu\nu}$ can be traceless or not.  

Remark that if, using the above results, one factors out the terms in $p$ and $k$ in Eq. (\ref{EQ-GeneralFormCoeff}) one obtains

\begin{equation}\label{EQ-GeneralFormCoeffInv}
\begin{split}
\overline{\square}\,\overline{h}_{\mu\nu} &+ a' \,\overline{R}\, \overline{h}_{\mu\nu} + 
b'\,\left(\overline{R}_{\mu\alpha}\,\overline{h}^\alpha_{~\nu}+\overline{R}_{\nu\alpha}\,\overline{h}^\alpha_{~\mu}\right)\\ 
&+ c'\overline{g}_{\mu\nu}\overline{R}_{\alpha\beta}\,\overline{h}^{\alpha\beta}
+ e \left(\overline{\nabla}_\mu\overline{\nabla}_\alpha\,\overline{h}^\alpha_{~\nu}
+ \overline{\nabla}_\nu\overline{\nabla}_\alpha\,\overline{h}^\alpha_{~\mu}\right)\\
& + p\, \overline{C}_{\mu\alpha\beta\nu}\overline{h}^{\alpha\beta}
+ f \,\overline{g}_{\mu\nu}\overline{\nabla}_\alpha\overline{\nabla}_\beta\,\overline{h}^{\alpha\beta}\\ 
&+ k\, \overline{g}_{\mu\nu} \left(\overline{\square} - \frac{1}{4}\frac{d-2}{d-1} \overline{R}\right)\overline{h}
+ l  \left( \overline{\nabla}_\mu \overline{\nabla}_\nu +\overline{\nabla}_\nu \overline{\nabla}_\mu \right) \overline{h}\\
&+ m'\, \overline{g}_{\mu\nu} \overline{R}\,\overline{h} + n'\, \overline{R}_{\mu\nu}\overline{h} = 0.
\end{split}
\end{equation}
In this expression, the prime denotes the coefficients in which the contributions of $p$ and $k$ have been extracted (that is $a' = a\vert_{p=0}$, {\it etc.}),
and $\overline{C}$ is the conformal transform of the Weyl tensor:
\begin{equation*}
\begin{split}
C_{\mu\alpha\beta\nu} &= R_{\mu\alpha\beta\nu} - \frac{1}{d - 2}\left(g_{\mu\beta} R_{\alpha\nu} - g_{\alpha\beta} R_{\mu\nu} \right.\\
                       &\left . + R_{\mu\beta} g_{\alpha\nu} - R_{\alpha\beta} g_{\mu\nu}\right)\\ 
                       &+ \frac{R}{(d-1)(d-2)} \left(g_{\mu\beta} g_{\alpha\nu} - g_{\alpha\beta} g_{\mu\nu} \right).
\end{split}
\end{equation*}
The two terms whose coefficients are $p$ and $k$ in Eq. (\ref{EQ-GeneralFormCoeffInv}) transform as
\begin{align*} 
&\overline{C}_{\mu\alpha\beta\nu} \overline{h}^{\alpha\beta} 
= \Omega^{(r-2)} C_{\mu\alpha\beta\nu}  h^{\alpha\beta},\\
&\overline{g}_{\mu\nu} \left(\overline{\square} - \frac{1}{4}\frac{d-2}{d-1} \overline{R}\right)\overline{h} = 
\Omega^{(r-2)} g_{\mu\nu} \left(\square - \frac{1}{4}\frac{d-2}{d-1} R\right) h,
\end{align*}
and thus do not alter the conformal invariance, which is consistent with the arbitrariness of $p$ and $k$. Note that since
for $d=3$ the Weyl tensor vanishes identically, the $p$ term carries no supplementary degree of freedom
and could be, in that case, dropped from Eq. (\ref{EQ-GeneralFormCoeffInv}).

We remark finally that  one can derive the conformal equation (\ref{EQ-GeneralFormCoeff}) from the following Lagrangian 
\begin{equation*}
\begin{split}
\mathcal{L} &= \sqrt{\vert g \vert} \left\{\frac{1}{2}  (\nabla_\alpha h_{\mu\nu})^2  - \frac{a}{2} R (h_{\mu\nu})^2 
- b R_\alpha^{~\beta} h^{\alpha\nu} h_{\nu\beta} 
   \right.\\
&- c R_{\alpha\beta} h^{\alpha\beta} h+  e \nabla_\alpha h^{\alpha\nu} \nabla^\beta h_{\nu\beta} + \frac{p}{2} R_{\mu\alpha\beta\nu} h^{\mu\beta} h^{\alpha\nu} 
 \\
& \left.+  f \nabla_\alpha h^{\alpha\beta} \nabla_\beta h+ \frac{k}{2} (\nabla_\alpha h)^2  -\frac{m}{2} R h^2 \right \}. 
\end{split}
\end{equation*}
We point out for future reference that setting $h=0$ in this Lagrangian before obtaining the Euler equations do not lead to the traceless 
version of Eq. (\ref{EQ-GeneralFormCoeff}).
\subsection*{Maximally symmetric spaces}
In a maximally symmetric space, the Riemann tensor and its contractions reduce to  
\begin{align}\label{EQ-RiemMaxSym}
& R_{\mu\alpha\beta\nu} = \kappa \left(g_{\mu\beta}g_{\alpha\nu} - g_{\mu\nu}g_{\alpha\beta}\right),\nonumber\\ 
& R_{\mu\nu} = (d-1) \kappa  g_{\mu\nu},\\ 
& R = d (d-1) \kappa, \nonumber
\end{align}
where $\kappa$ is a constant. If the space $(M, \overline{g})$ is a maximally symmetric space, the use of the above formulas together with the expression 
of the coefficients $a, b, \ldots,n$ allows us to recast the conformally invariant Eq. (\ref{EQ-GeneralFormCoeff}) in the form, 
\begin{equation}\label{EQ-CaseMaxSym}
\begin{split}
\overline{\square}\,\overline{h}_{\mu\nu} 
&- \frac{4}{d+2} \left(\overline{\nabla}_\mu\overline{\nabla}_\alpha\,\overline{h}^\alpha_{~\nu} + \overline{\nabla}_\nu\overline{\nabla}_\alpha\,
\overline{h}^\alpha_{~\mu}\right)\\&
+ \frac{8}{d(d+2)} \overline{g}_{\mu\nu}\overline{\nabla}_\alpha\overline{\nabla}_\beta\,\overline{h}^{\alpha\beta} 
 - \frac{d^2 - 2d +8}{4 d (d-1)}\, \overline{R}\, \overline{h}_{\mu\nu} \\
&+\, k \overline{g}_{\mu\nu} \left(\overline{\square} -  \frac{1}{4}\frac{(d-2)}{(d-1)}\overline{R} \right)\overline{h}\\
&+ \frac{4}{d(d+2)}  \left( \overline{\nabla}_\mu \overline{\nabla}_\nu +\overline{\nabla}_\nu \overline{\nabla}_\mu \right) \overline{h}\\
&+ \frac{8}{d^2(d+2)(d-1)} \overline{g}_{\mu\nu} \overline{R}\,\overline{h}= 0.
\end{split}
\end{equation}

As an example, the above equation for a traceless field in a four dimensional de Sitter background reads
\begin{equation*}
\begin{split}
\overline{\square}\,\overline{h}_{\mu\nu} 
&- \frac{2}{3} \left(\overline{\nabla}_\mu\overline{\nabla}_\alpha\,\overline{h}^\alpha_{~\nu} + \overline{\nabla}_\nu\overline{\nabla}_\alpha\,
\overline{h}^\alpha_{~\mu}\right)\\&
+ \frac{1}{3} \overline{g}_{\mu\nu}\overline{\nabla}_\alpha\overline{\nabla}_\beta\,\overline{h}^{\alpha\beta} 
 + 4 H^{2}\, \overline{h}_{\mu\nu} = 0,
\end{split}
\end{equation*}
with $H^2=-\overline{R}/12$.
This is the only conformally invariant linear wave equation on the de Sitter background for a traceless second rank tensor $h_{\mu\nu}$. 
This equation has already been obtained in \cite{DeserNepo} in a different context (the change from dS to AdS is obtained through the 
substitution $H^2\rightarrow -H^2$).

\section{The dimension $d=2$}\label{Sec-dim2} 
The $d=2$ case is very different since the classical geometrical tensors reduce to Eqs.(\ref{EQ-RiemMaxSym}) where 
$\kappa$ is now a scalar field. 
Under a conformal transformation, the relations (\ref{EQ-WeylRiemann}-\ref{EQ-WeylScalRicci}) still apply with $d=2$, and  the 
Eqs.(\ref{EQ-RiemMaxSym})
are also satisfied in the conformally related two-dimensional space. Now, the procedure explained in Sec. \ref{Sec-Equation} remains correct 
from equation (\ref{EQ-GeneralForm}) to equation (\ref{EQ-GeneralFormCoeff}). The system 
described in Appendix \ref{AP-DetailCalc} is also valid, however the details of its resolution are different. 
Solving this system 
and taking into account equations (\ref{EQ-RiemMaxSym}) leads to the $d=2$ conformal wave equation:
\begin{equation}
\begin{split}
\overline{\square}\,\overline{h}_{\mu\nu} 
&- \left(\overline{\nabla}_\mu\overline{\nabla}_\alpha\,\overline{h}^\alpha_{~\nu}
+ \overline{\nabla}_\nu\overline{\nabla}_\alpha\,\overline{h}^\alpha_{~\mu}\right) 
+ \overline{g}_{\mu\nu}\overline{\nabla}_\alpha\overline{\nabla}_\beta\,\overline{h}^{\alpha\beta} 
\\
&-\overline{R}\,\overline{h}_{\mu\nu}+ k \overline{g}_{\mu\nu} \overline{\square}\,\overline{h} 
+ \frac{1}{2} \left( \overline{\nabla}_\mu \overline{\nabla}_\nu +\overline{\nabla}_\nu \overline{\nabla}_\mu \right) \overline{h}\\
&+\frac{\overline{R}}{2}\overline{g}_{\mu\nu}\overline{h} = 0.
\end{split}
\end{equation}
In this case the conformal weight  $r$ of $h_{\mu\nu}$ is found to be equal to $2$.
As in $d$-dimension, one can choose $h_{\mu\nu}$ traceless or not, moreover the coefficient $k$ is still free. 
That corresponds to the fact that 
the term $\square h$ is, by itself, conformally invariant.

\section{Comparison with other works}\label{Sec-Compar}

Let us first consider the  Eq. (24) of the work of Nepomechie \cite{Nepomechie}. A straightforward calculation using the definitions of the quantities
$\widetilde{S}_{ab}$,  $\tilde{h}_{ab}$ and of the vielbein $e_\mu^{~a}$ shows that this equation is the same as Eq. (\ref{EQ-GeneralFormCoeffInv}) with 
$p=k=0$. In \cite{Nepomechie} this equation is shown to be invariant under infinitesimal transformations  which are the dilations, the special
conformal transformations and the Lorentz transformations, that is the so$(2,d)$ algebra of the conformal group where the translations has been excluded.
In addition, the equation is by construction invariant under the change of coordinates. Although the Eq. (24) of \cite{Nepomechie} is not obtained directly from
a Minkowskian Lagrangian (what can be easily done) its derivation (in particular the use of the gauge constraints Eq. (14) of \cite{Nepomechie}) is very 
close to a so called 
Ricci gauging which is known to imply the Weyl invariance \cite{IorioORaifSacWiese}. This is anyway confirmed by our result which shows in addition 
that the Eq. (6) is unique.

We turn now to the Lagrangian derived by Erdmenger and Osborn \cite{ErdmengerOsborn}, Eq.(14) of their work, for traceless symmetric tensor 
field of arbitrary rank. For a (traceless) second rank tensor field in $d$ dimensions the corresponding Euler equations reads   
\begin{equation}\label{EQ-ErdmengerOsborn}
\begin{split}
&\overline{\square}\,\overline{h}_{\mu\nu} + a' \,\overline{R}\, \overline{h}_{\mu\nu} + 
b'\,\left(\overline{R}_{\mu\alpha}\,\overline{h}^\alpha_{~\nu}+\overline{R}_{\nu\alpha}\,\overline{h}^\alpha_{~\mu}\right)\\ 
&
+ e \left(\overline{\nabla}_\mu\overline{\nabla}_\alpha\,\overline{h}^\alpha_{~\nu}
+ \overline{\nabla}_\nu\overline{\nabla}_\alpha\,\overline{h}^\alpha_{~\mu}\right)+ A\, \overline{C}_{\mu\alpha\beta\nu}\overline{h}^{\alpha\beta}
 = 0,
\end{split}
\end{equation}
where $A$ is the parameter defined in Eq. (12) of \cite{ErdmengerOsborn}, $a',b',e$ are defined below Eq. (\ref{EQ-GeneralFormCoeffInv}). 
This equation is close to the Eq. (\ref{EQ-GeneralFormCoeffInv}) for a traceless field with $p=A$. The two terms whose coefficients in 
Eq. (\ref{EQ-GeneralFormCoeffInv}) are  $c'$ and $f$ are missing.
Using formulas of Appendix \ref{AP-DetailWeylTransf}
an inspection of the transform of each term present in Eq. (\ref{EQ-ErdmengerOsborn})  show that the operator defined in the l.h.s. 
fails to be conformally invariant. 
This comes from the fact that the conformally invariant constraint $h = 0$ has been already imposed in the Lagrangian,  Eq.(14) of \cite{ErdmengerOsborn}. 
Indeed, the Eq. (\ref{EQ-GeneralFormCoeffInv}) {of the present paper} can be derived from the Lagrangian 
\begin{equation*}
\begin{split}
\mathcal{L} &= \sqrt{\vert g \vert} \left\{\frac{1}{2}  (\nabla_\alpha h_{\mu\nu})^2  - \frac{a'}{2} R (h_{\mu\nu})^2 
- b' R_\alpha^{~\beta} h^{\alpha\nu} h_{\nu\beta} 
   \right.\\
&- c' R_{\alpha\beta} h^{\alpha\beta} h+  e \nabla_\alpha h^{\alpha\nu} \nabla^\beta h_{\nu\beta} + \frac{p}{2} C_{\mu\alpha\beta\nu} h^{\mu\beta} h^{\alpha\nu} 
 \\
& \left.+  f \nabla_\alpha h^{\alpha\beta} \nabla_\beta h+ \frac{k}{2} (\nabla_\alpha h)^2 \right.\\
& \left. -\frac{1}{2}\left(m'- \frac{k(d-2)}{4(d-1}\right) R h^2 \right \}, 
\end{split}
\end{equation*} 
which gives, up to an irrelevant factor two, the Eq.(14) of \cite{ErdmengerOsborn} for $h=0$. It is apparent on this expression that 
the terms which are missing in Eq. (\ref{EQ-ErdmengerOsborn}) in order to be conformally invariant come from the two terms with coefficients $c'$ and $f$, 
both mixing the trace with $h_{\alpha\beta}$. These terms do contribute to Eq. (\ref{EQ-GeneralFormCoeffInv}), even for $h=0$.
This explains why the conformally invariant Lagrangian of Erdmenger and Osborn, supplemented with the conformally invariant condition $h=0$ leads
to the non-conformally invariant Eq. (\ref{EQ-ErdmengerOsborn}). In order to obtain conformally invariant Euler equations, the constraint has to be
imposed after the derivation of the equations, 
that is to say after the functional derivative of the action has been done. 
Now, taking the trace of Eq. (\ref{EQ-ErdmengerOsborn}) and using the expression
of the coefficients $b',c',e$ and $f$ leads to~: 
\begin{equation*}
c'\overline{g}_{\mu\nu}\overline{R}_{\alpha\beta}\,\overline{h}^{\alpha\beta} + 
f \,\overline{g}_{\mu\nu}\overline{\nabla}_\alpha\overline{\nabla}_\beta\,\overline{h}^{\alpha\beta} = 0.
\end{equation*}
A solution of Eq. (\ref{EQ-ErdmengerOsborn}) will always satisfy this equation. 
Thus the set of solutions of (\ref{EQ-ErdmengerOsborn})  is contained in the set of solutions 
of our conformally invariant operator of Eq.\nolinebreak(\ref{EQ-GeneralFormCoeffInv}).

Let us now compare our results  for $d=4$ with that of Grishchuk and Yudin \cite{GrishchukYudin1}. 
To this end, we first take into account the two constraints
\begin{equation}\label{EQ-CondGY-TT}
 \overline{h}=0,~~\overline{\nabla}_\alpha \overline{h}^\alpha_\mu = 0,
\end{equation}
into Eq. (\ref{EQ-GeneralFormCoeff}) which rewrites
\begin{equation}\label{EQ-FormTT}
\begin{split}
\overline{\square}\,\overline{h}_{\mu\nu} &+ a \overline{R} \overline{h}_{\mu\nu} + 
b\left(\overline{R}_{\mu\alpha}\,\overline{h}^\alpha_{~\nu}+\overline{R}_{\nu\alpha}\,\overline{h}^\alpha_{~\mu}\right)\\ 
&+ c\overline{g}_{\mu\nu}\overline{R}_{\alpha\beta}\,\overline{h}^{\alpha\beta} + p \overline{R}_{\mu\alpha\beta\nu}\overline{h}^{\alpha\beta} = 0.
\end{split}
\end{equation}
Following the remark about invariant terms in  Sec. \ref{Sec-Equation} we can add to this equation the term 
$(-p + \tilde{d}) \,\overline{C}_{\mu\alpha\beta\nu} \overline{h}^{\alpha\beta}$, $\tilde{d}$ being here the arbitrary parameter appearing 
in the Eq. (47) of \cite{GrishchukYudin1}.  We then recover this expression up to the coefficient of the term 
$\overline{g}_{\mu\nu}\overline{R}_{\alpha\beta}\,\overline{h}^{\alpha\beta}$ which is found to take on the value $1/2$  
in the present treatment. This 
difference comes from the fact that in \cite{GrishchukYudin1} it is assumed, in addition to the above 
conditions (\ref{EQ-CondGY-TT}) that $W_\alpha \overline{h}^\alpha_\mu = W_\alpha h^\alpha_\mu = 0$, where $W := \mathrm{d} \log \Omega^2$, 
in order to maintain the ``transversality condition'':
$\nabla_\alpha h^\alpha_\mu = 0$.  Indeed, the resolution of the system of equations  given in appendix \ref{AP-DetailCalc} shows that the coefficient
$c$ is determined through Eqs. (\ref{EQ-Syst-8}, \ref{EQ-Syst-9}). The equation (\ref{EQ-Syst-8}) precisely refers to the term 
$W_\alpha h^\alpha_\mu = 0,$ which appears in the conformal transformation of the Eq. (\ref{EQ-GeneralFormCoeff}) in the terms 
(\ref{EQ-WeylTerme-p}-\ref{EQ-WeylTerme-e}). Thus, this is the condition 
$W_\alpha h^\alpha_\mu = 0$ which allows the coefficient of  $\overline{g}_{\mu\nu}\overline{R}_{\alpha\beta}\,\overline{h}^{\alpha\beta}$ 
to be free in \cite{GrishchukYudin1}.

Finally, one can also compare our results for $d=4$ with the more recent work of Faci \cite{Faci} Eq. (44).  Note that, to this end,
one should recast the term 
$\overline{\nabla}_\mu \overline{\nabla}_\nu \overline{h}$ in the equation of \cite{Faci} under the symmetric form 
$(1/2) \left( \overline{\nabla}_\mu \overline{\nabla}_\nu +\overline{\nabla}_\nu \overline{\nabla}_\mu \right) \overline{h}$. 
These two expressions are equal as can be verified thanks to the formula (\ref{EQ-Commutateur}). Then, one almost recovers the 
result of \cite{Faci} by first adding to Eq. (\ref{EQ-GeneralFormCoeff}) 
the invariant term $\lambda \overline{C}_{\mu\alpha\beta\nu} \overline{h}^{\alpha\beta}$, $\lambda$ being 
the arbitrary parameter appearing in the Eq. (44) of \cite{Faci}, and then setting $p=-1$ and $k=-1/3$ in the coefficients listed above.
The discrepancies are the value of factors $e$ and $p$ equal respectively to $1/3$ and $1$ in \cite{Faci} versus $-2/3$ and $-1$ in the present work. 
Consequently, according to our Eq. (\ref{EQ-GeneralFormCoeffInv}), the operator appearing in Eq. (44) of \cite{Faci} is not conformally invariant. 

\section{Conclusion}\label{Sec-conl}

In this article we have obtained the conformally invariant linear wave operator for a free, symmetric, second rank tensor field 
in a d-dimensional ($d\geqslant 2$) metric manifold. This result extends in some respects (unicity, $d=2$, explicit Weyl invariance) 
the result already obtained by Nepomechie \cite{Nepomechie}. Giving this operator we hope to clarify the situation amongst  
contradictory results present in the literature. In this context we explain, in particular, 
why the conformally invariant Lagrangian of Erdmenger and Osborn\cite{ErdmengerOsborn}, supplemented with the conformally invariant condition $h=0$ leads
to a non-conformally invariant equation.

\section*{Acknowledgements}
The authors wish to thanks R. Nepomechie for providing us Ref \cite{Nepomechie} and constructive remarks, S. Deser for useful comments,  and 
J. Queva  for providing us Ref \cite{ErdmengerOsborn}.


\appendix
\section{Conventions and some formula}\label{AP-DefFrom}
Throughout this paper we use the following convention for the Riemann tensor
\begin{equation*}
R^\nu_{~~\alpha \mu \gamma} = \partial_\mu \Gamma^\nu_{~\gamma \alpha}
-\partial_\gamma \Gamma^\nu_{~\mu \alpha}
+ \Gamma^\rho_{~\gamma \alpha}
\Gamma^\nu_{~\mu \rho}
- \Gamma^\rho_{~\mu \alpha}
\Gamma^\nu_{~\gamma \rho}.
\end{equation*}
The Ricci tensor and Ricci scalar are: 
$R_{\mu\nu}\nolinebreak=\nolinebreak R^\alpha_{~~\mu \alpha \nu}$ and $R= R^\mu_{~\mu}$.
Under a rescaling of the metric $\overline{g}_{\mu\nu} = \Omega^2(x) g_{\mu\nu}$ one has, with $W_\mu:=\nabla_\mu\log\Omega^2$.
\begin{equation}\label{EQ-WeylRiemann}
\begin{split}
&\overline{R}^\mu_{~~\alpha\beta\nu} = R^\mu_{~~\alpha\beta\nu} 
-\frac{1}{2}\left(\delta^\mu_\beta\nabla_\nu W_\alpha -  \delta^\mu_\nu\nabla_\beta W_\alpha\right)\\
&+\frac{1}{2}\left(g_{\alpha\beta}\nabla_\nu W^\mu - g_{\alpha\nu}\nabla_\beta W^\mu\right)
-\frac{1}{4}W_\alpha\left(\delta^\mu_\nu W_\beta - \delta^\mu_\beta W_\nu\right)\\
&+\frac{1}{4}W^\mu\left(g_{\alpha\nu} W_\beta - g_{\alpha\beta} W_\nu\right)
+\frac{1}{4}W^2\left(g_{\alpha\beta} \delta^\mu_\nu - g_{\alpha\nu} \delta^\mu_\beta\right),
\end{split}
\end{equation}
\begin{equation}\label{EQ-WeylRicci}
\begin{split}
\overline{R}_{\mu\nu} &= R_{\mu\nu} - \frac{d-2}{2}\nabla_\nu W_\mu - \frac{1}{2} g_{\mu\nu} \nabla\cdot W + \frac{d-2}{4}W_\mu W_\nu\\
&-\frac{d-2}{4}g_{\mu\nu}W^2,
\end{split}
\end{equation}
\begin{equation}\label{EQ-WeylScalRicci}
\Omega^2\overline{R}= R - (d-1) \nabla\cdot W -\frac{1}{4}(d-1)(d-2)W^2.
\end{equation}
Note that the conformally invariant equation for a scalar (of weight $-(d-2)/2$) reads with the above convention
\begin{equation*}
 \left(\square - \frac{1}{4}\frac{d-2}{d-1} R\right) \phi = 0.
\end{equation*}

\section{Weyl transformation of the terms of Eq.~(\ref{EQ-GeneralFormCoeff})}\label{AP-DetailWeylTransf}
We give here the transformation rules in $d$ dimension for all the term which enter the Eq. (\ref{EQ-GeneralFormCoeff}). The conformal 
weight of $h_{\mu\nu}$ is $r$.

\begin{equation*}
\begin{split}
\Omega^{2-r}&\overline{ \square} \,\overline{h}_{\mu\nu}
= \square h_{\mu\nu} +\frac{2r+d-6}{2}W\cdot\nabla h_{\mu\nu}
\\
&+\frac{1}{2}\left[(\frac{r}{2}-1)(r+d-2)-r+1\right]W^2h_{\mu\nu}\\
& + 
(\frac{r}{2}-1) \nabla\cdot W h_{\mu\nu}
-(W_\mu\nabla_\alpha h^\alpha_\nu + W_\nu\nabla_\alpha h^\alpha_\mu)\\
&+W_\alpha(\nabla_\mu h_\nu^\alpha 
+ \nabla_\nu h_\mu^\alpha)
-\frac{d}{4} W_\alpha(W_\mu h^\alpha_\nu +W_\nu h^\alpha_\mu )\\
&
+\frac{1}{2}g_{\mu\nu}W_\alpha W_\beta h^{\alpha\beta}+\frac{1}{2}W_\mu W_\nu h.
\end{split}
\end{equation*}

\begin{equation*}
\begin{split}
\Omega^{2-r}&\overline{ R} \,\overline{h}_{\mu\nu}
= R h_{\mu\nu} -(d-1)\nabla\cdot W h_{\mu\nu}\\
&-\frac{1}{4}(d-1)(d-2)W^2h_{\mu\nu}.
\end{split}
\end{equation*}

\begin{equation*}
\begin{split}
\Omega^{2-r}&(\overline{R}_{\mu\alpha} \overline{h}^\alpha_\nu+\overline{R}_{\nu\alpha} 
\overline{h}^\alpha_\mu)\\&
= R_{\mu\alpha} h^\alpha_\nu+R_{\nu\alpha} h^\alpha_\mu \\&
-\frac{d-2}{2} \left(h^\alpha_\mu\nabla_\alpha W_\nu  +\, h^\alpha_\nu\nabla_\alpha W_\mu \right) \\&
+
\frac{d-2}{4}W_\alpha(W_\mu h^\alpha_\nu +W_\nu h^\alpha_\mu )\\&
- \nabla\cdot W h_{\mu\nu}-\frac{d-2}{2}W^2h_{\mu\nu}.
\end{split}
\end{equation*}

\begin{equation}\label{EQ-WeylTerme-p}
\begin{split}
\Omega^{2-r}&\overline{R}_{\mu\alpha\beta\nu} \overline{h}^{\alpha\beta}
= R_{\mu\alpha\beta\nu} h^{\alpha\beta}-\frac{1}{4}g_{\mu\nu}W_\alpha W_\beta h^{\alpha\beta}\\&+
\frac{1}{2}g_{\mu\nu}h^{\alpha\beta}\nabla_\alpha W_\beta -\frac{1}{4}W^2h_{\mu\nu}\\&
-\frac{1}{2} \left(h^\alpha_\mu\nabla_\alpha W_\nu  +\, h^\alpha_\nu\nabla_\alpha W_\mu \right) \\&
+\frac{1}{4}W_\alpha\left(W_\mu h^\alpha_\nu +W_\nu h^\alpha_\mu\right)\\&
-\frac{1}{4} h \left( W_\mu W_\nu  -2\nabla_\mu W_\nu  -g_{\mu\nu} W^2\right).  
\end{split}
\end{equation}

\begin{equation}\label{EQ-WeylTerme-c}
\begin{split}
\Omega^{2-r}&\overline{g}_{\mu\nu}\overline{R}_{\alpha\beta}\overline{h}^{\alpha\beta}=
g_{\mu\nu}R_{\alpha\beta}h^{\alpha\beta}-\frac{d-2}{2}g_{\mu\nu}h^{\alpha\beta}\nabla_\alpha W_\beta\\&
+ \frac{d-2}{4}g_{\mu\nu}W_\alpha W_\beta h^{\alpha\beta}-\frac{1}{2}g_{\mu\nu} h\nabla\cdot W\\&
-\frac{d-2}{4}g_{\mu\nu} h W^2.
\end{split}
\end{equation}

\begin{equation}\label{EQ-WeylTerme-f}
\begin{split}
\Omega^{2-r}&\overline{g}_{\mu\nu}\overline{\nabla}_\alpha\overline{\nabla}_\beta \overline{h}^{\alpha\beta}=
g_{\mu\nu}\nabla_\alpha\nabla_\beta h^{\alpha\beta}\\&
+(r+d-3)g_{\mu\nu}W_\alpha \nabla_\beta h^{\alpha\beta}\\&
+\frac{r+d-2}{2}\frac{r+d-4}{2}g_{\mu\nu}W_\alpha W_\beta h^{\alpha\beta}\\&
+\frac{r+d-2}{2}g_{\mu\nu}h^{\alpha\beta}\nabla_\alpha W_\beta \\&
-\frac{1}{2}g_{\mu\nu}W\cdot \nabla h-\frac{1}{2}g_{\mu\nu} h\nabla\cdot W\\&
-\frac{r+d-4}{4}g_{\mu\nu} h W^2.
\end{split}
\end{equation}

\begin{equation}\label{EQ-WeylTerme-e}
\begin{split}
\Omega^{2-r}&(\overline{\nabla}_\mu\overline{\nabla}_\alpha \overline{h}^\alpha_\nu + \overline{\nabla}_\nu
\overline{\nabla}_\alpha \overline{h}^\alpha_\mu)
=(\nabla_\mu\nabla_\alpha h^\alpha_\nu + \nabla_\nu\nabla_\alpha h^\alpha_\mu)\\&
+\frac{r-4}{2}\left(W_\mu\nabla_\alpha h^\alpha_\nu + W_\nu\nabla_\alpha h^\alpha_\mu\right)
+g_{\mu\nu}W_\alpha \nabla_\beta h^{\alpha\beta}\\&
+\frac{r+d-2}{2}W_\alpha(\nabla_\mu h_\nu^\alpha + \nabla_\nu h_\mu^\alpha)\\&
+\frac{r+d-2}{2}\frac{r-4}{2}W_\alpha(W_\mu h^\alpha_\nu +W_\nu h^\alpha_\mu )\\&
+\frac{r+d-2}{2}\left(h^\alpha_\mu\nabla_\alpha W_\nu + h^\alpha_\nu\nabla_\alpha W_\mu \right.\\&
\left. +g_{\mu\nu}W_\alpha W_\beta h^{\alpha\beta}\right)\\&
-\frac{1}{2}(W_\mu \nabla_\nu + W_\nu \nabla_\mu)h-h\nabla_\mu W_\nu\\&
-\frac{1}{2}g_{\mu\nu} h W^2 - \frac{r-4}{2}W_\mu W_\nu h.
\end{split}
\end{equation}

\begin{equation*}
\begin{split}
\Omega^{2-r}&(\overline{\nabla}_\mu\overline{\nabla}_\nu + \overline{\nabla}_\mu\overline{\nabla}_\nu)
\overline{h}=(\nabla_\mu\nabla_\nu + \nabla_\mu\nabla_\nu)h\\&
+ (r-3)(W_\mu \nabla_\nu + W_\nu \nabla_\mu)h +(r-2)h\nabla_\mu W_\nu\\&
+ \frac{(r-2)(r-4)}{2}W_\mu W_\nu h + g_{\mu\nu}W\cdot \nabla h\\&
+ \frac{r-2}{2}g_{\mu\nu} h W^2.
\end{split}
\end{equation*}

\begin{equation*}
\begin{split}
\Omega^{2-r}&\overline{g}_{\mu\nu}\overline{\square} \,\overline{h} = g_{\mu\nu}\square h +
\frac{2r+d-6}{2}g_{\mu\nu}W\cdot \nabla h\\&
+\frac{1}{2}\left[(\frac{r}{2}-1)(r+d-2)-r+2\right]g_{\mu\nu} h W^2\\&
+(\frac{r}{2}-1)g_{\mu\nu} h\nabla\cdot W.
\end{split}
\end{equation*}

\begin{equation*}
\begin{split}
\Omega^{2-r}&\overline{g}_{\mu\nu}\overline{R}\,\overline{h} =g_{\mu\nu}Rh -(d-1)g_{\mu\nu} h\nabla\cdot W\\&
-\frac{(d-1)(d-2)}{4}g_{\mu\nu} h W^2.
\end{split}
\end{equation*}

\begin{equation*}
\begin{split}
\Omega^{2-r}&\overline{R}_{\mu\nu}\overline{h}=R_{\mu\nu}h-\frac{d-2}{2} h\nabla_\mu W_\nu
-\frac{1}{2}g_{\mu\nu} h\nabla\cdot W\\&
+\frac{d-2}{4}W_\mu W_\nu h-\frac{d-2}{4}g_{\mu\nu} h W^2.
\end{split}
\end{equation*}

\section{Determination of the coefficients of Eq.~(\ref{EQ-GeneralFormCoeff})}\label{AP-DetailCalc}
The system of equations for the coefficients of Eq. (\ref{EQ-GeneralFormCoeff}) to ensure conformal invariance reads
\begin{align}
\frac{1}{2}\left(\left(\frac{r}{2}-1\right)(r+d-2)-r+1\right)- \frac{a}{4}(d-1)(d-2)&\nonumber\\
-b\frac{d-2}{2}-\frac{p}{4} &=  0\label{EQ-Syst-1}\\
\left(\frac{r}{2}-1\right)-a(d-1)-b &= 0 \label{EQ-Syst-2}\\
1+e \frac{r+d-2}{2}&=0\label{EQ-Syst-3}\\
-1 + e\frac{r-4}{2}&= 0\label{EQ-Syst-4}\\
-\frac{d}{4}+\frac{1}{4} b (d-2)+\frac{p}{4}+\frac{1}{4} e\text{  }(r+d-2)(r-4)&=0\label{EQ-Syst-5}
\end{align}
\begin{align}
-\frac{1}{2} b (d-2)-\frac{p}{2}+\frac{1}{2} e (r+d-2)&=0\label{EQ-Syst-6}\\
\frac{2r +d-6}{2}&=0\label{EQ-Syst-7}\\
\frac{1}{2}-\frac{p}{4}+ c \frac{d-2}{4}+\frac{e}{2} (r+d-2) &\nonumber\\
+f\frac{r +d-2}{2}\frac{r +d-4}{2}&=0\label{EQ-Syst-8}\\
\frac{p}{2}- c \frac{d-2}{2}+f\frac{r +d-2}{2}&=0\label{EQ-Syst-9}\\
e+f(r+d-3)&=0\label{EQ-Syst-10}
\end{align}
\begin{align}
\frac{1}{2}-\frac{p}{4}-\frac{1}{2}(r-4)e+\frac{1}{2}(r-2)(r-4)l+\frac{1}{4}(d-2)n&=0\label{EQ-Syst-11}\\
\frac{p}{2}-e+(r-2)l-\frac{1}{2}(d-2)n&=0\label{EQ-Syst-12}\\
\frac{p}{4}-\frac{1}{4}(d-2)c-\frac{1}{2}e-\frac{1}{4}(r+d-4)f+\frac{1}{2}(r-2)l&\nonumber\\
+\frac{1}{2}\left(\left(\frac{r}{2}-1\right)(r+d-2)-r+2\right)k&\nonumber\\
-\frac{1}{4}(d-1)(d-2)m-\frac{1}{4}(d-2)n &=0\label{EQ-Syst-13}
\end{align}
\begin{align}
-\frac{1}{2}c -\frac{1}{2}f + \left(\frac{r}{2}-1\right)k - (d-1)m -\frac{1}{2}n &=0\label{EQ-Syst-14}\\
-\frac{1}{2}e +(r-3)l & =0\label{EQ-Syst-15}\\
-\frac{1}{2}f +l + \frac{1}{2}(2r +d-6)k & =0.\label{EQ-Syst-16}
 \end{align}
The equations (\ref{EQ-Syst-1}) to (\ref{EQ-Syst-4}), can be solved to give $a,b,e$ and $r$ as functions of $d$ and $p$. The 
relations (\ref{EQ-Syst-6}) to (\ref{EQ-Syst-7}) are then satisfied. The equations (\ref{EQ-Syst-8}-\ref{EQ-Syst-9}) can then be solved
to give $f$ as a function of $d$ and  $c$ as a function of $d$ and $p$. The remaining equations (\ref{EQ-Syst-11}-\ref{EQ-Syst-16}) are concerned by the terms of
Eq.(\ref{EQ-GeneralFormCoeff}) containing the trace $h$. From Eq. (\ref{EQ-Syst-15}) one obtains $l$ as a function of $d$, Eqs. (\ref{EQ-Syst-11}) 
then gives $n$ as function of $d$ and $p$. The Eq. (\ref{EQ-Syst-12}) is then always satisfied. The Eqs. (\ref{EQ-Syst-13}-\ref{EQ-Syst-14}) are 
found to be identical, one can then solve  (\ref{EQ-Syst-13}) to obtain $m$ as a function of $d$, $p$ and $k$ which remain free. The second term 
of the last equation (\ref{EQ-Syst-16}) vanishes identically and the remaining relation between $f$ and $l$ is satisfied. The solutions are collected in 
Sec. \ref{Sec-Equation}.


\begin{thebibliography}{AAA} \baselineskip=10pt
\bibitem{kastrup} H.~A~Kastrup, Ann. Phys. (Berlin) {\bf17}, 631 (2008).
\bibitem{wald} R.~Wald, \emph{General Relativity}, (The University of Chicago Press, 1984).
\bibitem{DeserNepo} S.~Deser and R.~Nepomechie, Phys. Lett. {\bf 132B}, 321 (1983).
\bibitem{GrishchukYudin1} L.~P.~Grishchuk and V.~M.~Yudin, J. Math. Phys. (N. Y.) {\bf 21}, 1168 (1980).
\bibitem{Grishchuk1} L.~P.~Grishchuk, Sov. Phys. JETP {\bf 40}, 409 (1975).
\bibitem{Parker} L.~Parker,  Phys. Rev. {\bf 183}, 1057 (1969).
\bibitem{Faci} S.~Faci, Classical Quantum Gravity, {\bf 30}, 115005 (2013).  
\bibitem{Nepomechie} R.~Nepomechie, Phys. Lett. {\bf 136B}, 33 (1984).
\bibitem{ErdmengerOsborn} J.~Erdmenger and H.~Osborn, Classical Quantum Gravity, {\bf 15}, 273 (1998).
\bibitem{IorioORaifSacWiese} A.~Iorio, L.~O'Raifeartaigh, I. Sachs and C.~Wiesendanger, Nucl. Phys. B, {\bf 495}, 433 (1997).



\end{thebibliography}
\end{document}